\documentstyle[12pt]{article}

\parskip=\medskipamount

\begin{document}


\vspace{8mm}

\begin{center}

{\Large \bf Phase and Scaling Properties of Determinants Arising in
Topological Field Theories\footnote{Work supported by FORBAIRT
Scientific Research Program SC/94/218}} \\

\vspace{12mm}

{\large David H. Adams\footnote{Supported by Trinity College Postgraduate
Award, FORBAIRT Basic Research Award}
and Siddhartha Sen}

\vspace{4mm}

School of Mathematics, Trinity College, Dublin 2, Ireland. \\

\vspace{1ex}

email: dadams,sen@maths.tcd.ie \\

\end{center}

\begin{abstract}

In topological field theory determinants of maps with negative as well as
positive eigenvalues arise. We give a generalisation of the
zeta-regularisation technique to derive expressions for the phase and
scaling-dependence of these determinants. For theories on odd-dimensional
manifolds a simple formula for the scaling dependence is obtained in terms
of the dimensions of cohomology spaces. This enables a non-perturbative
feature of Chern-Simons gauge theory to be reproduced by semiclassical
methods

\end{abstract}

\newpage

Topological field theories (TFTs) are of interest because they
provide examples of quantum field theories which are exactly solvable
and because they provide a new way of looking at topological
invariants of manifolds \cite{S(LMP)},\cite{W(TFT)}. A  particular TFT,
the Chern-Simons gauge theory on 3-dimensional manifolds, has led to new
invariants \cite{S(Baku)},\cite{W(Jones)}. (For a review of TFTs see
\cite{BirRak}).

In topological field theory, given a topological action functional
$S(\omega)$ for fields $\omega$ on a manifold $M\,$, an object of interest
is the partition function
\begin{equation}
Z(\beta)=\int_{\Gamma}{\cal D\/}\omega\,e^{-{\beta}S(\omega)}
\label{1}
\end{equation}
where the formal integration is over the infinite-dimensional vectorspace
$\Gamma$ of fields $\omega\,$. We have included in (\ref{1}) a scaling
parameter $\beta$ which we allow to be complex-valued. (Typically $\beta$
is either real or purely imaginary; it is often taken to be a constant
equal to 1 or $-i\,$). For the cases we consider in this paper the manifold
$M$ is required to be compact, without boundary and oriented (e.g. a sphere
of arbitrary dimension).

For a wide class of TFTs where the action $S(\omega)$ is quadratic
(see (\ref{17}) below for a specific example)
the partition function can be formally
evaluated by the method of A.~Schwarz \cite{S(LMP)},\cite{S(NuclP)}. This
leads to an expression for (\ref{1}) consisting of a product of determinants
of certain maps associated with $S(\omega)\,$. One of these determinants
is\footnote{This should really be $det(\beta\frac{1}{\pi}
\widetilde{T})^{-1/2}$ since $\int_{-\infty}^{\infty}e^{-x^2}\,dx
=\sqrt{\pi}\,$. However the numerical factor $1/\pi$ in the determinant
is usually considered to be irrelevant and discarded.}
\begin{equation}
det(\beta\widetilde{T})^{-1/2}
\label{2}
\end{equation}
where $\widetilde{T}$ is obtained by discarding the zero-modes of the
selfadjoint map $T$ on $\Gamma$ given by
\begin{equation}
S(\omega)=<\omega\,,T\omega>\;.
\label{3}
\end{equation}
The inner product $<\cdot\,,\cdot>$ in $\Gamma$ used to obtain $T$ from
$S(\omega)$ in (\ref{3}) is constructed from a Euclidean metric on $M$
(as in \cite[p.437]{S(NuclP)}). The other determinants in the expression for
the partition function appear because of the zero-modes of $T\,$. They are
all real-valued and do not involve the parameter $\beta\,$. Hence the
phase of the partition function (\ref{1}) and its dependence on the
scaling parameter $\beta$ are determined solely by the determinant (\ref{2}).
The determinants in the expression for the partition function are determinants
of maps on infinite-dimensional vectorspaces and must therefore be regularised
in order to obtain a finite expression.
This is done using the zeta-regularisation technique.

In this paper we consider a subtlety in the zeta-regularisation of the
determinant (\ref{2}).
The zeta-regularisation technique requires the map to be positive, i.e. all
its eigenvalues must be positive. But the action functional $S(\omega)$
of the TFT typically takes negative as well as positive values, so from
(\ref{3}) it follows that $\widetilde{T}$ has negative as well as positive
eigenvalues. This problem was sidestepped by Schwarz (and in most subsequent
work on TFTs) by replacing $\widetilde{T}$ in (\ref{2}) by the positive
map $|\widetilde{T}|\,$.
This map is defined in the following way: Take a basis $\lbrace\omega_j
\rbrace$ for $\Gamma$ of eigenvectors of $T$ with eigenvalues
$\lbrace\lambda_j\rbrace\,$, then $|T|$ is defined by setting
$|T|\omega_j=|\lambda_j|\omega_j$ and $|\widetilde{T}|$ is obtained from
$|T|$ by discarding the zero-modes.

For a particular case (with $\beta=i\,$) E.~Witten has shown in
\cite[\S2]{W(Jones)} how the zeta-regularisation technique can be generalised
to evaluate (\ref{2}) (see also \cite[\S7.2]{At}).
He found that a complex phase factor appears,
determined by $\eta(0\,;\,T)\,$, where $\eta(s\,;\,T)$ is the eta-function
of $T\,$. In this paper we evaluate the determinant (\ref{2}) (with
arbitrary $\beta\in{\bf C}\,$) for all the above-mentioned TFTs considered
by Schwarz. This is done using a straightforward generalisation of the
usual zeta-regularisation technique and analytic continuation in $\beta\,$,
and generalises the calculation of Witten mentioned above. The following
expression is obtained: Let ${\bf C_+}$ and ${\bf C_-}$ denote the upper
and lower halfplanes of {\bf C} respectively, then for $\beta=|\beta|
e^{i\theta}\in{\bf C_{\pm}}$ with $\theta\in[-\pi\,,\pi]$ we find
\begin{equation}
det(\beta\widetilde{T})^{-1/2}
=e^{-\frac{i\pi}{4}((\frac{2\theta}{\pi}\mp1)\zeta\;
\pm\;\eta)}\,|\beta|^{-\zeta/2}\,det(|\widetilde{T}|)^{-1/2}
\label{4}
\end{equation}
where $\zeta$ and $\eta$ are the analytic continuations to $s=0$ of the
zeta-function $\zeta(s\,;\,|T|)$ and eta-function $\eta(s\,;\,T)$
respectively (defined as in (\ref{8}) and (\ref{11}) below) and
$det(|\widetilde{T}|)^{-1/2}$ is defined by the usual zeta-regularisation
technique. In particular, for $\lambda\in{\bf R_+}$ we get
\begin{equation}
det(\lambda\widetilde{T})^{-1/2}=e^{\pm\frac{i\pi}{4}(\zeta-\eta)}\,
\lambda^{-\zeta/2}\,det(|\widetilde{T}|)^{-1/2}
\label{5}
\end{equation}
and
\begin{equation}
det(i\lambda\widetilde{T})^{-1/2}=e^{-\frac{i\pi}{4}\eta}\,\lambda^{-\zeta/2}
\,det(|\widetilde{T}|)^{-1/2}\;.
\label{6}
\end{equation}
Note that for $\beta\in{\bf R}$ there is a phase ambiguity in (\ref{4})
(analogous to the ambiguity in $\sqrt{-1}={\pm}i\,$) while there is no
ambiguity for $\beta\in{\bf C}-{\bf R}\,$ (e.g. when $\beta$ is purely
imaginary).
It is not immediately obvious that the phase and scaling factors in
(\ref{4})--(\ref{6}) are finite, since this requires the zeta-function
$\zeta(s\,;\,|T|)$ and eta-function $\eta(s\,;\,T)$ to have analytic
continuations regular at $s=0\,$. If $T$ were elliptic then this would
follow from standard results in mathematics; however for the cases arising
in TFTs the map $T$ in (\ref{3}) is {\it not} elliptic. We will
nevertheless show below that
$\zeta(s\,;\,|T|)$ and $\eta(s\,;\,T)$ do in fact have analytic continuations
regular at $s=0\,$, so the expressions (\ref{4})--(\ref{6}) are finite.
(We do not claim that this is a  new mathematical result, but for the sake
of completeness we give a simple derivation).
We also derive a simple formula for $\zeta=\zeta(0\,;\,|T|)$ in terms of
the dimensions of certain cohomology
spaces when $M$ has odd dimension ((\ref{24}) below).
This leads to a simple expression for the scaling dependence of
(\ref{4})--(\ref{6}).

Determinants of the form (\ref{2}) are also relevant for TFTs where the
action $S(\omega)$ contains higher order terms as well as the quadratic term.
In this case determinants of the form (\ref{2}) appear in the semiclassical
approximation for the partition function of the theory. A particular TFT
with non-quadratic action is the Chern-Simons gauge theory on
3-dimensional manifolds (given by (\ref{28}) below), which was shown to be
solvable by E.~Witten in \cite{W(Jones)}. We will discuss below how the
dependence of the semiclassical approximation on the parameter $k$ in this
theory can be obtained from our calculation of (\ref{2}). Because it is
a solvable theory for a field with self-interactions the Chern-Simons
gauge theory provides a ``mathematical laboratory'' in which predictions
of perturbation theory can be tested. A basic prediction of
perturbative quantum field theory
is that the semiclassical approximation should coincide
with the non-perturbative expression for the partition function in the
limit where the parameter $k$ of the theory becomes large. The large $k$
limit of the partition function, with gauge group $SU(2)\,$,
has been explicitly calculated by Witten's
non-perturbative method for a large number of 3-manifolds in a program
initiated by D.~Freed and R.~Gompf \cite{FG(PRL)},\cite{FG(CMP)}. They
found that the $k-$dependence of the partition function in this limit
is given by a simple expression ((\ref{33}) below). Subsequent work by
L.~Jeffrey \cite{J} and L.~Rozansky \cite{Roz} has verified this expression
for large classes of 3-manifolds. The expression we obtain below for the
$k-$dependence of the semiclassical approximation turns out to be identical
to this non-perturbative expression. Thus we reproduce a non-perturbative
feature of the Chern-Simons gauge theory from perturbation theory.

Before evaluating (\ref{2}) we briefly recall the usual zeta-regularisation
technique. The zeta-function of a positive selfadjoint linear map $A$ is
defined by
\begin{equation}
\zeta(s\,;\,A)=\sum_j\frac{1}{\lambda_j^s}{\qquad}\;\;\;\;\;\;s\in{\bf C}
\label{8}
\end{equation}
where $\lbrace\lambda_j\rbrace$ are the non-zero eigenvalues of $A\,$
(so $\lambda_j>0$ for all $\lambda_j$ in (\ref{8})) with each eigenvalue
appearing the same number of times as its multiplicity. With $\widetilde{A}$
obtained from $A$ by discarding the zero-modes we can formally write
\begin{equation}
det(\widetilde{A})=\prod_j\lambda_j=e^{-\zeta'(0\,;\,A)}\,.
\label{9}
\end{equation}
When $A$ acts on an infinite-dimensional vectorspace $\zeta(s\,;\,A)$ is
divergent around $s=0\,$. However in many cases of interest it turns out that
$\zeta(s\,;\,A)$ is well-defined for $Re(s)>>0$ and extends by analytic
continuation to a meremorphic function on ${\bf C}$ which is regular at
$s=0\,$. Then we can use the analytic continuation of $\zeta(s\,;\,A)$ to
give well-defined meaning to the r.h.s. of (\ref{9}) and use this to define
$det(\widetilde{A})$ in (\ref{9}). For $\beta\in{\bf R_+}$ we then obtain
a well-defined expression for $det(\beta\widetilde{A})$ by replacing
$\widetilde{A}$ by $\beta\widetilde{A}$ in (\ref{9}). This leads to
\begin{equation}
det(\beta\widetilde{A})=\beta^{\zeta(0\,;\,A)}\,e^{-\zeta'(0\,;\,A)}\,.
\label{10}
\end{equation}
Using (\ref{10}) we can define $det(\beta\widetilde{A})$ for arbitrary
$\beta\in{\bf C}$ via analytic continuation in $\beta\,$. To do this we
must fix a convention for defining $z^a$ for $z\in{\bf C}$ and
$a\in{\bf R}\,$. The natural way to do this is to write $z=|z|e^{i\theta}$
with $\theta\in[-\pi\,,\pi]$ and set $z^a=|z|^ae^{i{\theta}a}\,$. This is
well-defined for all $a\in{\bf R}$ provided $z\not\in{\bf R_-}\,$; if
$z\in{\bf R_-}$ then there is a phase ambiguity. With this convention
(\ref{10}) is defined for all $\beta\in{\bf C}$ up to a phase ambiguity
for $\beta\in{\bf R_-}\,$.
Finally, recall that the
eta-function of a selfadjoint linear map $B\,$ (which may have both positive
and negative eigenvalues) is defined by
\begin{equation}
\eta(s\,;\,B)=\sum_k\frac{1}{(\lambda_k^{(+)})^s}-\sum_l\frac{1}
{(-\lambda_l^{(-)})^s}
\label{11}
\end{equation}
where $\lbrace\lambda_k^{(+)}\rbrace$ and $\lbrace\lambda_l^{(-)}\rbrace$ are
the strictly positive- and strictly negative eigenvalues of $B$ respectively.
In many cases of interest it turns out that $\eta(s\,;\,B)$ is well-defined
for $Re(s)>>0$ and extends by analytic continuation to a meremorphic
function on ${\bf C}$ which is regular at $s=0\,$.

We shall now evaluate the determinant (\ref{2}). Formally we have
\begin{equation}
det(\beta\widetilde{T})^{-1/2}=(\,det({\beta}T_+)\,det({\beta}T_-)\,)^{-1/2}
\label{12}
\end{equation}
where $T_+$ and $T_-$ are obtained from $T$ by restricting to the strictly
positive- and strictly negative modes respectively. Note that $-T_-$ is
positive (i.e. has positive eigenvalues) and that
\begin{eqnarray}
\zeta(s\,;\,|T|)&=&\zeta(s\,;\,T_+)+\zeta(s\,;\,-T_-) \label{13} \\
\eta(s\,;\,T)&=&\zeta(s\,;\,T_+)-\zeta(s\,;\,-T_-) \label{14}
\end{eqnarray}
{}From (\ref{12}), using (\ref{9}), (\ref{10}) and (\ref{13}) we get
\begin{eqnarray}
det(\beta\widetilde{T})^{-1/2}&=&det({\beta}T_+)^{-1/2}\,det((-\beta)\,
(-T_-))^{-1/2} \nonumber \\
&=&\beta^{-\zeta(0\,;\,T_+)/2}\,(-\beta)^{-\zeta(0\,;\,-T_-)/2}
\,e^{(\zeta'(0\,;\,T_+)+\zeta'(0\,;\,-T_-))/2} \nonumber \\
&=&\beta^{-\zeta(0\,;\,T_+)/2}\,(-\beta)^{-\zeta(0\,;\,-T_-)/2}\,
det(|\widetilde{T}|)^{-1/2} \label{15}
\end{eqnarray}
For $\beta=|\beta|e^{i\theta}\in{\bf C_{\pm}}$ with $\theta\in[-\pi\,,\pi]$
we have $-\beta=|\beta|e^{i(\theta\mp\pi)}$ with $\theta\mp\pi\in[-\pi\,,\pi]$
and a simple calculation using (\ref{13}) and (\ref{14}) shows
\begin{equation}
\beta^{-\zeta(0\,;\,T_+)/2}\,(-\beta)^{-\zeta(0\,;\,-T_-)/2}=
e^{-\frac{i\pi}{4}((\frac{2\theta}{\pi}\mp1)\zeta(0\,;\,|T|)\;\pm\;
\eta(0\,;\,T))}\,.
\label{16}
\end{equation}
Substituting this in (\ref{15}) gives (\ref{4}).

As pointed out previously, for the expression (\ref{4}) to have well-defined
meaning $\zeta(s\,;\,|T|)$ and $\eta(s\,;\,T)$ must be regular at $s=0\,$.
We will now show that this is the case for the cases of interest in TFT.
In doing so we derive a simple formula for $\zeta(0\,;\,|T|)$ when $M$
has odd dimension. For the sake of concreteness we will work with a
specific topological action functional
\begin{equation}
S(\omega)=\int_M\omega{\wedge}d_m\omega\,.
\label{17}
\end{equation}
The fields $\omega$ are the real-valued differential forms
on $M$ of degree $m$ and $d_q$ denotes the exterior derivative on
$q-$forms. $M$ is required to have odd dimension $n=2m+1$ and we assume that
$m$ is odd, since for $m$ even (\ref{17}) is
identically zero. The quadratic action functionals in other TFTs are
generalisations of (\ref{17}) and it is easily checked that the following
arguments continue to hold for these. A choice of metric on $M$ enables us to
construct an inner product in the space of differential forms in the usual
way (as in \cite[p.437]{S(NuclP)}) and with this we can write
\begin{equation}
S(\omega)=<\omega\,,T\omega>\;\;\;,\;\;\;\;T=*d_m
\label{18}
\end{equation}
where $*$ is the Hodge star-map (as in \cite[p.437]{S(NuclP)}). We denote
the space of $q-$forms on $M$ by $\Omega^q(M)$ and define the
Laplace-operator on $\Omega^q(M)$ by
\begin{equation}
\Delta_q=d_q^*d_q+d_{q-1}d_{q-1}^*\;\;\;,\;\;\;\;q=0,1,\dots,n
\label{19}
\end{equation}
(with $d_{-1}=d_n=0\,$). We will derive a relationship between the
zeta-function of $|T|$ and the zeta-functions of $\Delta_0\,,\,\Delta_1\,,
\dots,\,\Delta_m\,$. To do this we will use the following simple
observation: Consider linear maps $A$ and $B$ on a vectorspace, satisfying
$AB=BA=0\,$. Then if $\lbrace\lambda_j\rbrace$ denotes the collection of
non-zero eigenvalues of $A+B$ (with each eigenvalue appearing the same
number of times as its multiplicity) we have
\begin{equation}
\lbrace\lambda_j\rbrace=\lbrace\lambda_k'\rbrace\,\cup\,\lbrace\lambda_l''
\rbrace
\label{20}
\end{equation}
where $\lbrace\lambda_k'\rbrace$ and $\lbrace\lambda_l''\rbrace$ are the
non-zero eigenvalues of $A$ and $B$ respectively. (This is an elementary
fact in linear algebra which is easily verified). Setting
$A=d_q^*d_q$ and $B=d_{q-1}d_{q-1}^*$ the property $AB=BA=0$ follows
from $d_qd_{q-1}=0\,$, and it follows from (\ref{19}) and (\ref{20}) that
\begin{eqnarray}
\zeta(s\,;\,\Delta_q)&=&\zeta(s\,;\,d_q^*d_q)+\zeta(s\,;\,d_{q-1}d_{q-1}^*)
\nonumber \\
&=&\zeta(s\,;\,d_q^*d_q)+\zeta(s\,;\,d_{q-1}^*d_{q-1})
\label {21}
\end{eqnarray}
where we have used the simple fact that for any linear map $C$ the maps
$C^*C$ and $CC^*$ have the same non-zero eigenvalues. A simple induction
argument based on (\ref{21}) and starting with $\zeta(s\,;\,d_m^*d_m)=
\zeta(s\,;\,\Delta_m)-\zeta(s\,;\,d_{m-1}^*d_{m-1})$ shows that
\begin{equation}
\zeta(s\,;\,d_m^*d_m)=(-1)^m\sum_{q=0}^m(-1)^q\zeta(s\,;\,\Delta_q)\,.
\label{22}
\end{equation}
The map $T$ in (\ref{18}) has the property $T^2=d_m^*d_m$ and from the
definition (\ref{8}) we see that $\zeta(s\,;\,T^2)=\zeta(2s\,;\,|T|)\,$.
It follows from (\ref{22}) that
\begin{equation}
\zeta(s\,;\,|T|)=(-1)^m\sum_{q=0}^m(-1)^q\zeta(\frac{s}{2}\,;\,\Delta_q)\,.
\label{23}
\end{equation}
This shows that $\zeta(s\,;\,|T|)$ is well-defined for $Re(s)>>0$ with
analytic continuation regular at $s=0\,$, since the zeta-functions of the
Laplace-operators $\Delta_q$ are known to have these properties
(see e.g. \cite[ch.28]{S(QFT)}).
When $dimM$ is odd we have $\zeta(0\,;\,\Delta_q)=-dimH^q(d)$
(see \cite[ch.28]{S(QFT)}), where $H^q(d)=ker(d_q)\,\Big/\,Im(d_{q-1})$ is
the q'th cohomology space of $d\,$. It follows from (\ref{23}) that in
this case
\begin{equation}
\zeta(0\,;\,|T|)=(-1)^{m+1}\sum_{q=0}^m(-1)^qdimH^q(d)\,.
\label{24}
\end{equation}

We now consider the eta-function $\eta(s\,;\,T)\,$. A standard result in
elliptic operator theory states that the
eta-function of an elliptic selfadjoint
map is regular at $s=0\,$. (This is due to M.~Atiyah, V.~Patodi and I.~Singer
\cite{APS3} in the case where $dimM$ is odd, and P.~Gilkey \cite{G} when
$dimM$ is even). The map $T$ in (\ref{18}) is selfadjoint but not elliptic.
However we can construct an elliptic selfadjoint map $D$ such that
$\eta(s\,;\,D)=\eta(s\,;\,T)\,$, from which it follows that $\eta(s\,;\,T)$
is regular at $s=0\,$. For $q=0,1,\dots,m$ we extend $d_q$ to a map on
$\oplus_{q=0}^m\Omega^q(M)$ by setting $d_q=0$ on $\Omega^p(M)$ for
$p{\ne}q\,$. We define the map $\widetilde{D}$ on $\oplus_{q=0}^m\Omega^q(M)$
by $\widetilde{D}=\sum_{q=0}^m(d_q+d_q^*)$ and set $D=T+\widetilde{D}\,$, with
$T$ as in (\ref{18}). $D$ is clearly selfadjoint and a simple calculation
using the property $d_qd_{q-1}=0$ shows that $D^2=\sum_{q=0}^m\Delta_q\,$,
which is elliptic, so $D$ is elliptic. It is immediate from the definitions
of $\widetilde{D}$ and $T$ that $T\widetilde{D}=\widetilde{D}T=0$ and it
follows from (\ref{20}) that
\begin{equation}
\eta(s\,;\,D)=\eta(s\,;\,T)+\eta(s\,;\,\widetilde{D})\,.
\label{25}
\end{equation}
To show $\eta(s\,;\,D)=\eta(s\,;\,T)$ we must show that $\eta(s\,;\,
\widetilde{D})=0\,$. We consider the eigenvalue equation $\widetilde{D}
\omega=\lambda\omega$ with $\omega=\oplus_{q=0}^m\omega_q
\in\oplus_{q=0}^m
\Omega^q(M)\,$. This is equivalent to the collection of equations
\begin{equation}
d_q\omega_q+d_{q+1}^*\omega_{q+2}=\lambda\omega_{q+1}\;\;,\;\;
q=0,1,\dots,m-1
\label{26}
\end{equation}
(with $\omega_{m+1}=0\,$). If $\omega$ is a solution to (\ref{26}) then
we set $\omega'=\oplus_{q=0}^m\omega'_q$ with $\omega'_q=(-1)^q\omega_q\,$.
Then
\begin{equation}
d_q\omega'_q+d_{q+1}^*\omega'_{q+2}=(-1)^q(d_q\omega_q+d_{q+1}^*
\omega_{q+2})=(-1)^q\lambda\omega_{q+1}=-\lambda\omega'_{q+1}
\end{equation}
and it follows from (\ref{26}) that $\widetilde{D}\omega'=-\lambda\omega'\,$.
This shows that there is a one-to-one correspondence
$\omega\leftrightarrow\omega'$ between eigenvectors for $\widetilde{D}$ with
eigenvalue $\lambda$ and eigenvectors with eigenvalue $-\lambda\,$,
and it follows from the definition (\ref{11}) that
$\eta(s\,;\,\widetilde{D})=0$ as claimed.
(The statement $\eta(s\,;\,T)=\eta(s\,;\,D)$
is similar to \cite[proposition(4$\cdot$20)]{APS1}).

Finally, as promised, we apply our results to
the semiclassical approximation for the partition
function of the Chern-Simons gauge theory on 3-manifolds.
The partition function of this theory is
\begin{equation}
Z(k)=\int{\cal D\/}A\,e^{ikS(A)}\;\;\;\;\;\;\;\;,\;\;\;\;\;\;\;\;\;k\in{\bf Z}
\label{27}
\end{equation}
where
\begin{equation}
S(A)=\frac{1}{4\pi}\int_MTr(A{\wedge}dA+\frac{2}{3}A{\wedge}A{\wedge}A)\,.
\label{28}
\end{equation}
The gauge fields $A$ are the $1-$forms on $M$ with values in the Lie
algebra of the gauge group $SU(N)\,$. The parameter $k$ is required to be
integer-valued, then the integrand in (\ref{27}) is gauge-invariant.
An expression for the semiclassical approximation for (\ref{27}) can be
obtained from the invariant integration method of A.~Schwarz
\cite[\S5]{S(CMP)}. (We emphasise that Schwarz's method is ideally suited
for evaluating the semiclassical approximation for (\ref{27}).
This method leads to the appearance of inverse volume factors
$V(H_A)^{-1}$ in the integrand of the expression ((\ref{29}) below) for the
semiclassical approximation (see \cite[\S5, formula(1)]{S(CMP)}), where
$H_A$ is the subgroup of gauge transformations which leaves the gauge field
$A$ unchanged.
These factors are necessary to reproduce the
numerical factors appearing in the large $k$ limit of the non-perturbative
expression for the partition function
and have not been obtained in a self-contained way in other evaluations of
the semiclassical approximation for
the Chern-Simons partition function\footnote{These volume factors were
put in by hand in the expression for the semiclassical approximation given
by L.~Rozansky in \cite{Roz} and shown to lead to agreement with the large
$k$ limit of the non-perturbative expression for the partition function
for large classes of 3-manifolds}. We will be discussing this in more detail
in a future paper; see also \cite{AdSe(hep-th)}.)
The expression obtained from Schwarz's method for the semiclassical
approximation for (\ref{27}) has the form
\begin{equation}
Z_{sc}(k)=\int_{{\cal M\/}^F}{\cal D\/}A\,e^{ikS(A)}\mu(k;A)
\label{29}
\end{equation}
where ${\cal M\/}^F$ is the moduli space of flat gauge fields
modulo gauge transformations. (The flat gauge fields are the solutions to
the field equations corresponding to (\ref{28})). The integrand
$e^{ikS(A)}\mu(k;A)$
is gauge-invariant and is therefore a well-defined function
on ${\cal M\/}^F\,$. The quantity $\mu(k;A)$
is given by \cite[\S5, formula(1)]{S(CMP)}
and its dependence on $k$ enters through the determinant
\begin{equation}
det(ick\widetilde{T}_A)^{-1/2}\;\;\;,\;\;\;\;T_A=*d_1^A
\label{30}
\end{equation}
where $c$ is a numerical constant (involving $\pi\,$)
and $d_q^A$ is the flat covariant derivative on the Lie algebra-valued
$q-$forms obtained from $d_q$ by ``twisting'' by the flat gauge field $A\,$.
(See \cite[\S15.2]{S(TFP)}
for the definition of this). The results above concerning
the map $T$ in (\ref{18}) generalise for the map $T_A$ in (\ref{30}).
Since in the present case $dimM=3\,$, $m=1$ and it follows from (\ref{6})
and (\ref{24}) that the $k-$dependence of the determinant in (\ref{30})
is given by
\begin{equation}
k^{-\zeta(0\,;\,|*d_1^A|)/2}=k^{(-dimH^0(d^A)+dimH^1(d^A))/2}\,.
\label{31}
\end{equation}
It follows that in the limit of large $k$ the $k-$dependence of the
semiclassical approximation (\ref{29}) (ignoring phase factors)
is given by
\begin{equation}
k^{\Bigl(\stackrel{max}{A}\lbrace-dimH^0(d^A)/2+dimH^1(d^A)/2\rbrace\Bigr)}
\label{33}
\end{equation}
where the maximum is taken over the flat gauge fields. This is precisely the
$k$ dependence \cite[formula(1.37)]{FG(CMP)} of the large $k$ limit of
the partition function (\ref{27}) obtained from non-perturbative
calculations\footnote{In \cite{Roz} L.~Rozansky gave a heuristic argument
for why it was plausible that the $k-$dependence of the integrand in
(\ref{29}) should be given by (\ref{31}). This did not involve calculating
the $k-$dependence of the determinant in (\ref{30}).}.

We illustrate this with a specific example. When $M$ is the 3-sphere
the expression for the partition function obtained from Witten's
non-perturbative method \cite[\S4]{W(Jones)} with gauge group $SU(2)$ is
\begin{equation}
Z(k)=\sqrt{\frac{2}{k+2}}\,sin\biggl(\frac{\pi}{k+2}\biggr)\;\;\sim\;\;
\sqrt{2}{\pi}k^{-3/2}\;\;\;\;\mbox{for}\;\;k\to\infty\,.
\label{32}
\end{equation}
Since $\pi_1(S^3)$ is trivial the only flat gauge field on the 3-sphere
up to gauge equivalence is the trivial field $A\!=\!0\,$, and in this case
we have $dimH^0(d^A)\!=\!dim(su(2))\,dimH^0(S^3)$
$=\!3$ and $dimH^1(d^A)=dim(su(2))\,dimH^1(S^3)
=0\,$. It follows from (\ref{31}) that the $k-$dependence
of the semiclassical approximation in this case is $\;\sim\,k^{-3/2}\,$,
in agreement with the large $k$ limit of (\ref{32}).

\vspace{1ex}

{\it Acknowledgement.} We thank A.~Schwarz
for drawing our attention to \cite{S(NuclP)} and \cite{S(Baku)} and for
helpful comments.

\end{document}